# Preparation of atomically clean and flat Si(100) surfaces by low-energy ion sputtering and low-temperature annealing

J. C. Kim [a,b], J.-Y. Ji [b], J. S. Kline [a], J. R. Tucker [a], and T.-C. Shen [b,*]

[a] Department of Electrical and Computer Engineering, University of Illinois, Urbana, IL 61801
[b] Department of Physics, Utah State University, Logan, UT 84322

[*] Corresponding author. Department of Physics, Utah State University, UT 84322, USA. Tel: 435-797-7852; FAX: 435-797-2492; E-mail: tcshen@cc.usu.edu

**Abstract**

Si(100) surfaces were prepared by wet-chemical etching followed by 0.3-1.5keV Ar ion sputtering, either at elevated or room temperature. After a brief anneal under ultrahigh vacuum conditions, the resulting surfaces were examined by scanning tunneling microscopy. We find that wet-chemical etching alone cannot produce a clean and flat Si(100) surface. However, subsequent 300eV Ar ion sputtering at room temperature followed by a 700°C anneal yields atomically clean and flat Si(100) surfaces suitable for nanoscale device fabrication.

Atomically clean and well-ordered Si(100) surfaces can be routinely obtained by flashing the samples to 1250°C [1]. However, in nanoscale device fabrication, concerns over alterations in the dopant profile and damage to prefabricated structures limit the thermal budget. Therefore, low temperature preparation techniques such as *ex situ* wet chemical cleaning and *in situ* ion sputtering are appealing. Etching in $NH_4F$ solution creates a very flat Si(111) monohydride surface [2], and etching in diluted HF solution produces somewhat ordered Si(100) surfaces [3]. However, it is very difficult to eliminate contamination on the *ex situ* treated surfaces. Oxygen and carbon have long been recognized as the two major contaminants on Si surfaces, and carbon is by far the most difficult to remove. Based on Auger Electron Spectroscopy (AES), Joyce reported that impurity-free surfaces can be prepared by 0.2-2 keV inert gas ion bombardment followed by annealing at ~700°C [4]. However, besides impurities, surface smoothness is also a major concern in nanoscale device fabrication. The irradiation damage, including displacement of silicon atoms and incorporation of irradiating ions and surface contaminants into the bulk, must also be addressed. Effects of ion energy, species, flux, fluence, and substrate temperatures during ion sputtering and annealing have all been subject to intensive studies, but a consistent picture is still lacking due to the very wide range of ion sputtering and sample preparation conditions, and limited sensitivity of the characterization instruments [5-13].

The ion-surface interaction as a function of ion energy and substrate temperature on clean Si(100) surfaces has been previously investigated. At low ion fluences and sufficiently high substrate temperatures, Si(100) surfaces remain crystalline [5-8].

Bedrossian observed that the Si(100) surface sputtered by 200eV Xe ions at 500°C is smoother than the surface sputtered by 200eV ions at room temperature (RT) followed by annealing at 500°C [8], suggesting that ion sputtering at elevated temperatures is favorable in obtaining smooth surfaces.

The scenario becomes complicated, however, by the presence of surface contamination from wet-chemical etching and residual gas adsorption in ultrahigh vacuum (UHV) chambers. Based on the results of medium energy ion scattering (MEIS) Al-Bayati *et al*. concluded that the thickness of the disordered layer due to Ar ion sputtering decreases with substrate temperatures from RT to 400°C at ion energy 510eV, and increases with ion energies from 60eV to 510eV at 200°C [9]. Furthermore, even low energy (110eV) Ar ion bombardment on native-oxide-covered or H-terminated Si(100) surfaces at RT can create a significant amount of the damage in the near-surface region which cannot be annealed out at 800°C [10]. Ar ion sputtering also creates more stable defects at 800°C than at RT [10]. which is consistent with an earlier observation by Bean *et al*.[11]. The subsurface damage is attributed to the formation of SiC, due to dynamic ion mixing and incorporation of Ar [10], which can be enhanced by elevated temperatures during ion sputtering [12]. Note that subsurface damage may not be revealed by some surface techniques, as evidenced by relatively good reflection high energy electron diffraction (RHEED) patterns obtained in sputtering of native-oxide-covered Si(100) surfaces with 100-200eV Ar ions at substrate temperatures of 400-500°C [13]. Thus, real space measurement of the surface topography could shed light on some of these issues.

In this letter, we focus on two issues relevant to nanoscale device fabrication: (1) how to remove surface contaminants with low energy ion sputtering, and (2) how to minimize the irradiation damage in the near-surface region. In general, wet-chemically etched surfaces are contaminated with carbon. We found that complete amorphization of wet-chemically etched Si(100) surfaces by 300eV Ar ion sputtering at RT, followed by a ~5min 700°C anneal, can result in a flat, atomically clean surface with vacancy lines. Higher energy ion sputtering and ion sputtering at elevated temperatures causes multilayer ion erosion.

Our samples were cut from both n- and p-type 0.1Ω-cm Si(100) wafers. Ion sputtering was performed in the UHV chamber backfilled with Ar to $5 \times 10^{-5}$ Torr. Typical ion beam current is 11μA, which corresponds to an estimated ion flux of $3 \times 10^{13}$ $cm^{-2}s^{-1}$. The ion incidence angle is 50° from surface normal. The details of H-termination and patterning that we use have been reported previously [14].

Despite extensive investigations [3,15-17], wet-chemical etching of Si(100) has yet to produce large-scale atomically flat surfaces. Even after the standard RCA[18] or Shiraki[19] cleaning procedures and prompt loading into the UHV system, we cannot remove the surface carbon contamination completely [20]. Typically after RCA cleaning and 3% HF etching, the Si(100) surfaces are characterized by small hillocks [16]. If the surface is free from contamination, a 3-min anneal at 500°C should produce a very flat surface, based on our observations of silicon homoepitaxy. As depicted in Fig.1(a), the surface annealed at 500°C has irregular terraces with a roughness of 1.4nm. Fig 1(b) shows SiC crystallites on top of wide and flat terraces after annealing the chemically-etched surface at above 700°C. This result demonstrates that, in a typical laboratory environment, great effort is required to minimize carbon contamination on Si surfaces from water, ambient air or carbon-containing molecules inside the UHV chamber.

Since ion bombardment can induce SiC formation at RT at energies as low as 100eV [13], Si surface morphology during ion sputtering in the presence of carbon is a result of competition between sputtering removal and creation of SiC. We observe that sputtering of HF-etched Si surfaces by 300-400eV Ar ions at RT with a fluence of $1 \times 10^{17}$ cm$^{-2}$ completely amorphizes the near-surface region with a negligible amount of carbon contamination. Subsequent annealing at 500°C for 5min yields a Si(100) surface with a number of small terraces of lateral dimension < 20 nm and a roughness of more than 10 atomic layers [20]. However, after annealing at 700°C for 5min the surface is atomically clean and flat, as shown in Fig. 2(a). After the annealing, the surface has been terminated by H *in situ* to protect the surface from contaminants; small bright protrusions are a few remaining dangling bond sites. Fig. 2(b) shows that this surface can be selectively patterned by desorbing H atoms using 7eV electrons emitted from the STM tip [14]. It is not clear at this point if the line vacancies running perpendicular to the underlying dimer rows in Fig. 2(b) are induced by Ar retention or vacancy diffusion [21]. However, we believe that no significant adverse effects would result from these vacancy lines in subsequent nanoscale device fabrication.

Despite a higher sputtering yield to remove surface contaminants, many undesirable effects occur in ion bombardment at elevated temperatures, including higher diffusion of contaminants and incorporated Ar, enhanced formation of subsurface defect complexes such as SiC, dislocation, and vacancy clusters [10,12], and surface roughening [22]. Ion sputtering of an HF-etched Si(100) surface by 400eV Ar ions at 700°C yields a surface with many SiC crystallites [20]. Although surface carbon contamination can be completely eliminated by 1.5keV Ar ion sputtering at 700°C, terraces up to 12 atomic layers are produced [20]. Therefore, even when Si(100) surfaces prepared by ion sputtering at elevated temperatures display a decent RHEED pattern [13], those surfaces may not be flat enough for nanoscale device fabrication and could have substantial subsurface carbon that has not yet agglomerated into SiC crystallites.

We also studied surface roughening after ion sputtering on clean Si(100) surfaces at elevated temperatures. At 700°C, 400eV Ar ion sputtering results in a very flat surface; but raising the ion energy to 600eV and 800eV leads to increasing surface roughening as shown in Fig. 3. This result illustrates the interplay of competing kinetic processes during ion sputtering at elevated temperatures, including the production of surface defects (vacancies and adatoms) and their diffusion and annihilation at step edges [22].

RT ion sputtering produces an amorphized layer near the surface and suppresses the mobility of irradiation damages. With sufficient sputtering rate to minimize carbon concentration in the amorphized layer, it is possible to regenerate a clean and flat surface by a minimal annealing. For a dopant concentration of $1 \times 10^{20}$ cm$^{-3}$, As, P, and Sb diffuse less than 0.4nm and B diffuses ~ 3.2nm in 5 min at 700°C [23]. acceptable for nanoscale device template applications.

This work was supported by NSF under Grant No. 9875129 (TCS), ARDA/ARO under Grant No. DAAD19-00-R-0007, and DARPA QuIST under Grant No. DAAD19-01-1-0324.

**Figure Captions**

Fig. 1: STM image of Si(100) surface prepared by RCA cleaning followed by (a) 3% HF etching and 3 min annealing at 500°C and (b) 5 min annealing at 1000°C only.

Fig. 2: (a) STM image of H-terminated Si(100) surface prepared by 300eV Ar ion sputtering at a fluence $1\times10^{17}$ cm$^{-2}$ and subsequent 5min annealing at 700°C. (b) Surface of (a) after desorbing a line of hydrogen by 7V electrons from the STM tip.

Fig. 3: STM image of Si(100) surface after (a) 800eV Ar ion sputtering with a fluence $5\times10^{15}$ cm$^{-2}$ at 600°C and (b) 600eV Ar ion sputtering with a fluence of $4\times10^{15}$ cm$^{-2}$ at 700°C.

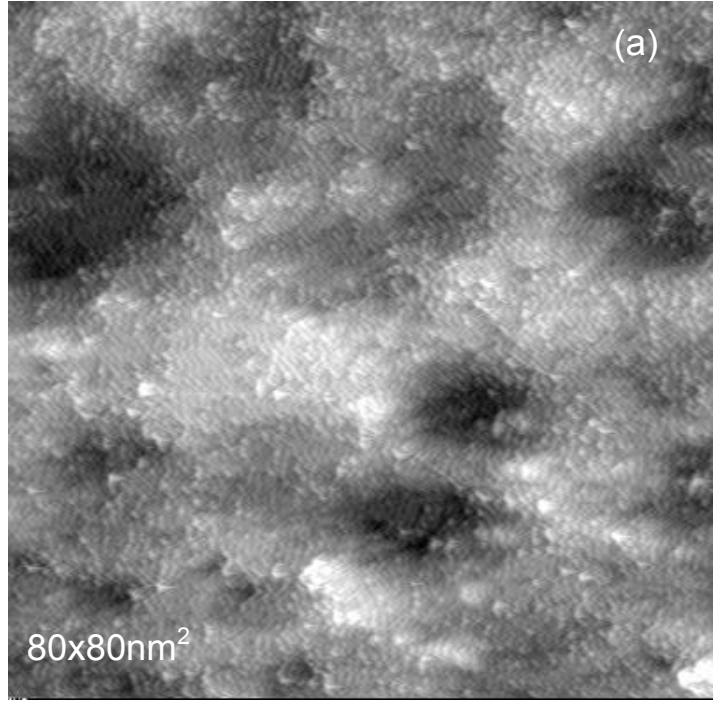

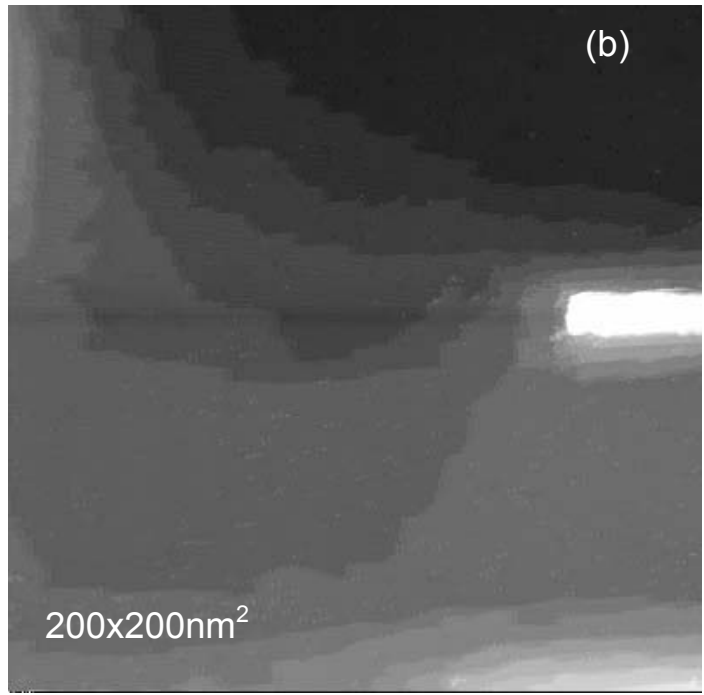

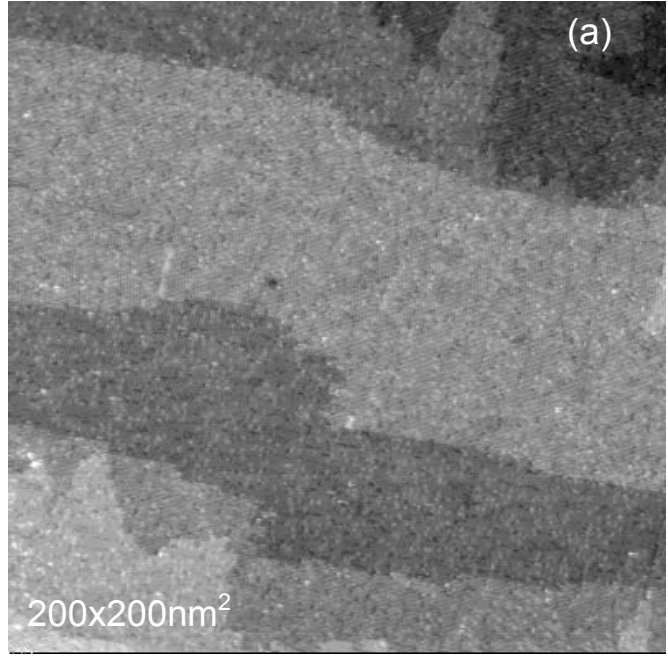

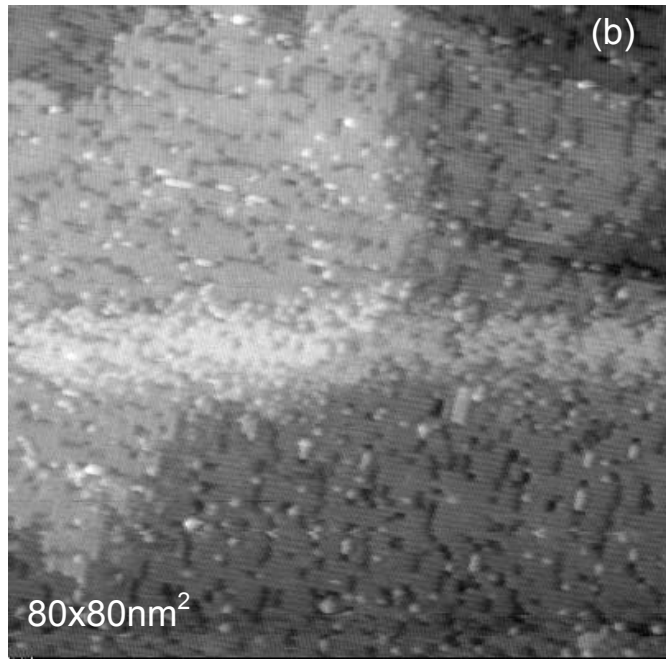

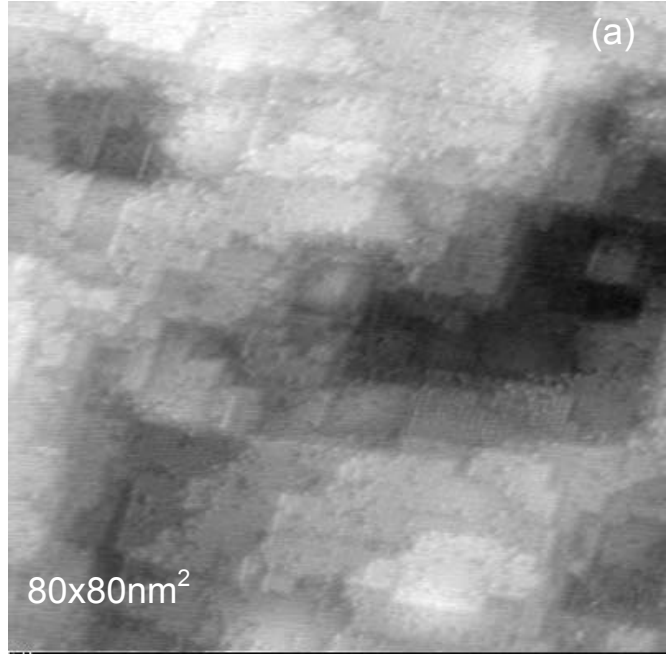

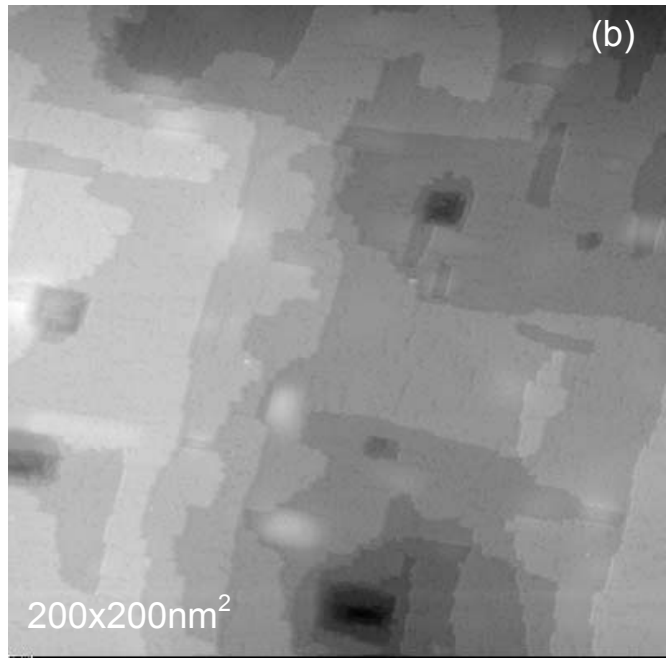